\documentclass[letter]{aa} 

\usepackage{graphicx}
\usepackage{txfonts}
\usepackage{subfig,float,siunitx}

\usepackage{amstext} 
\usepackage{array}   
\newcolumntype{L}{>{$}l<{$}} 
\begin{document}

   \title{Connections between the cycle-to-cycle light curve and O$-$C
variations of the Blazhko RR Lyrae stars}
   \titlerunning{C2C and O-C variations on Blazhko RR Lyrae stars}

   \author{J\'ozsef M. Benk\H{o}
   }

   \institute{Konkoly Observatory, HUN-REN Research Centre for Astronomy and Earth Sciences,
              MTA Centre of Excellence, \\ Konkoly Thege u. 15-17., 1121 Budapest, Hungary, 
              \email{benko@konkoly.hu}
             }

   \date{Received Month Day, Year; accepted Month Day, Year}


\abstract
   {Recent studies have shown that the irregular O$-$C variations observed in many non-Blazhko RR Lyrae stars may result from random, cycle-to-cycle (C2C) variations in their light curves. However, centuries-long data series reveal that the O$-$C diagrams of Blazhko stars exhibit particularly large-amplitude, irregular variations.}
   {In this Letter, we extend the previous investigation of non-Blazhko stars to \textit{Kepler} Blazhko stars to explore the role of C2C variations in the O$-$C diagrams.}
   {We derived the O$-$C diagrams from \textit{Kepler} space telescope light curves using a precise template-fitting method. Based on their Fourier analyses, we also constructed residual O$-$C diagrams that were pre-whitened for frequencies associated with the Blazhko effect. We then fitted the same statistical models to both types of O$-$Cs that we had previously applied to non-Blazhko stars.
   }
   {The optimal statistical model includes the C2C variation for 74\% of the O$-$C curves in our Blazhko sample, and the parameter describing the strength of the C2C variation is significantly larger than that obtained for non-Blazhko stars. This may explain the strong irregular O$-$C variations previously observed in Blazhko stars. Furthermore, we found a strong positive correlation between the C2C variation strength and the amplitude of the frequency-modulation component of the Blazhko effect, indicating a connection between the two phenomena.
   } 
   {}

   \maketitle
%

\section{Introduction}

RR Lyrae stars are horizontal-branch stars located within the classical instability strip of the Hertzsprung–Russell diagram. They pulsate radially in the fundamental (RRab), first-overtone (RRc), or both modes (RRd) \citep[see e.g.][]{Catelan_2009}. In addition to the brightness variations caused by radial pulsation, the light curves of many RR Lyrae stars display simultaneous amplitude (AM) and frequency (FM) modulations with periods much longer than the pulsation period ranging from several days to several years (see cf. \citealt{Jurcsik_Smitola2016, Netzel2018, Skarka2020}). This phenomenon is known as the Blazhko effect (for detailed reviews, see, e.g. \citealt{Smolec_Blazhko} and \citealt{Kovacs_Blazhko2016}).

Numerous studies have investigated long-term period variation of RR Lyrae stars using O$-$C diagrams (e.g. \citealt{LeBorgne2007,Szeidl2011, Jurcsik2012, Prudil2019, Hajdu2021}). These analyses have revealed that some RR Lyrae stars exhibit slow, regular period changes consistent with stellar evolutionary models, whereas others show much faster, irregular variations. Although this latter group includes several RRab and RRc stars that do not show the Blazhko effect, Blazhko stars are always found among them. Their irregular O$-$C variations generally exhibit higher amplitudes than those of non-Blazhko stars. In other words, it seems that the Blazhko effect gives rise to irregular O$-$C variations. However, it remains unclear how this occurs and why the amplitudes of irregular O$-$C variations are larger in Blazhko than in non-Blazhko stars.

\citet{Benko_2025} recently demonstrated that the vast majority of irregular O$-$C diagrams of non-Blazhko RR Lyrae stars observed by the \textit{Kepler} and \textit{TESS} space telescopes can be explained by random, cycle-to-cycle (C2C) variations in their light curves. In this Letter, we extend that analysis to Blazhko stars, in order to examine whether similar variations can account for their O$-$C behaviour as well.

\section{Data and methods}

Since the characteristic period of the Blazhko effect is typically much longer than the pulsation period, we used the four-year \textit{Kepler} data set \citep{Borucki2010}, which provides uninterrupted, high-precision space photometry suitable for detailed C2C analysis.
\citet{Benko2014} investigated the Blazhko RR Lyrae stars observed by the \textit{Kepler} space telescope, and in this study we used the corrected (tailor-made) light curves provided by them. We used 15 Blazhko RRab stars from \citet{Benko2014}.  
The work of \citet{Forro2022}, who identified ten Blazhko RR Lyrae stars in the background pixels of the original \textit{Kepler} field, proved useful in compiling a larger, homogeneous sample, observed with the same instrument, cadence, and photometric setup.
We performed O$-$C analyses for all the RR Lyrae stars found by \citet{Forro2022}, two of them appear to be new Blazhko candidates. Thus, our final sample consists of 27 RR Lyrae (25 RRab and 2 RRc) stars. Table~\ref{table:main} lists the stars included in this study. 

For consistency with previous work on non-Blazhko stars \citep{Benko_2025}, we constructed the O$-$C diagrams using a template-fitting method \citep{Hertzsprung1919}. For the definition and a detailed discussion of O$-$C diagrams, see \citealt{Sterken2005}. The used template fitting method is described in more detail in \citet{Benko2019}. In brief:
we derived a template from the Fourier fit of the average light curve, fitted Fourier sums with different numbers of terms to the folded light curve, and determined the optimal number of terms using the Akaike information criterion. For our sample, this number ranged from 4 to 39. The template was then shifted horizontally and scaled vertically by a unique multiplier of each pulsation cycle. The phase shift obtained from the fit was adopted as the O$-$C value. The resulting diagrams are shown in red in Figs.~\ref{fig:o-c}.  

For high-amplitude O$-$C diagrams, the template-fitting method and the classical polynomial fitting yield similar results, but for low-amplitude cases the template-fitting method reveals considerably more detail and visibly better precision. The improved accuracy of the method is illustrated by V838 Cyg, whose small-amplitude Blazhko modulation nevertheless allowed a reliable O$-$C curve to be derived, in contrast to \citet{Benko2014}.
\citet{Benko2014} reported an average timing uncertainty of about one minute ($\approx 6.9\cdot10^{-4}$~d). In this work, we estimated individual alignment uncertainties using Monte Carlo simulations; these are shown as light yellow error bars in Fig.~\ref{fig:o-c}. 
Typical 1$\sigma$ uncertainties of individual O$-$C points range from $3.4\times10^{-5}-1.7\times10^{-3}$~d for the original \textit{Kepler} targets and from $9\times10^{-4}-4.7\times10^{-3}$~d for the background stars.

The blue curves in Fig.~\ref{fig:o-c} present the first O$-$C diagrams for the stars from \citet{Forro2022}. In the best cases, the accuracy of these diagrams matches that of the original \textit{Kepler} targets (red curves). The higher dispersion of some curves is partly due to the faintness of these background stars and, in some quarters, to partial flux loss caused by imperfect photometric masks.  

The O$-$C diagrams were Fourier-analysed using the {\sc Period04} program \citep{Lenz2005}. The highest peak was assigned to the main Blazhko frequency. \citet{Benko2014} found that, in multi-periodic Blazhko stars, the relative strengths of the AM and FM components may differ — one period can dominate the amplitude modulation while another dominates the frequency modulation. The Fourier spectra of the O$-$C curves contain only the FM component of the Blazhko effect, therefore, the Blazhko frequency referred to here as dominant actually means dominant in FM. These frequencies are listed in the fourth column of Table~\ref{table:main}. During the Fourier analysis of the O$-$C diagrams, we found that the significant frequencies correspond to the Blazhko frequencies themselves, their harmonics, sub-harmonics, and — in cases of multiple Blazhko modulations — linear combinations of those frequencies. Pre-whitening all such terms (typically one to eight per star) yielded the O$-$C residuals. Although the light-curve shape variations of Blazhko stars may introduce systematic phase shifts in the determination of O$-$Cs, any such effects occur on the Blazhko timescale and therefore appear in the O$-$C spectra; consequently, the residuals remain free of these systematics. The frequencies removed, along with their possible identifications, are given in Table~\ref{table:all_freq}.

We also detected significant frequencies associated with long-term trends, the origins of which remain uncertain. They could arise from instrumental effects, very long Blazhko cycles, or additional physical processes; alternatively, they might reflect apparent period changes caused by C2C variations, as proposed in our previous work. Because the focus of this paper is the study of such variations, we did not pre-whiten these frequencies — they therefore remain in what we refer to as the “O$-$C residuals.” No residual curve was constructed for the two RRc stars, as their O$-$C variations are extremely long-period (if periodic at all) and cannot be clearly separated from the Blazhko or other long-term effects.  

\citet{Benko_2025} analysed the O$-$C diagrams of \textit{Kepler} non-Blazhko RR Lyrae stars using the statistical framework of \citet{Koen2006}. In this work, we adopt the same approach, assuming that the instantaneous pulsation period $P_i$ can be expressed as the sum of three components:
\begin{equation}
    P_i = \langle P \rangle + \sum_{j=1}^i \xi_j + \eta_i ,
\end{equation}
where $\langle P\rangle$ is the constant mean period, the summation term represents a general, smooth period variation, $\xi_j$ is assumed to be a zero-mean, uncorrelated random variable, and $\eta_i$ accounts for the effect of C2C variations in the light curve on the instantaneous period.  

Moreover, the construction of O$-$C diagrams is inherently affected by timing noise (or phase noise):
\begin{equation}
    t_i = T_i + e_i ,
\end{equation}
where $t_i$ is the measured time, $T_i$ the true time, and $e_i$ a random variable with zero mean. The phase noise $\phi_i$ and timing noise $e_i$ are related through $\phi_i = 2\pi e_i / P$, where $P$ is the pulsation period. Thus, three random variables are involved — $e$, $\xi$, and $\eta$ — with corresponding standard deviations $\sigma_e$, $\sigma_\xi$, and $\sigma_\eta$.  

\citet{Benko_2025} investigated four statistical models based on these assumptions to determine which provided the best fit to the observational data. The four models are as follows: M1, phase noise only; M2, phase noise plus C2C variation; M3, a combination of phase noise and real period change; and M4, including all three effects simultaneously.

\section{Results and discussion}
\subsection{New Blazhko candidates}

\citet{Forro2022} identified ten Blazhko stars in the background pixels of the \textit{Kepler} space telescope but did not examine their O$-$C diagrams. 
\citet{Forro2022} noted that KPP23 and KPP26 show amplitude changes, which they attributed to instrumental effects. Although the light curve of KPP23 is of relatively poor quality, its O$-$C diagram clearly shows variability (see Fig.~\ref{fig:o-c}). The corresponding Fourier spectrum contains two significant frequencies: $f_{\mathrm B}=0.00543\pm0.000046$~d$^{-1}$ (S/N = 5.99) and $2f_{\mathrm B}=0.01086\pm0.000046$~d$^{-1}$ (S/N = 12.16). The derived period, $P_{\mathrm B}=184.1\pm1.6$ d, is indeed close to twice the duration of a \textit{Kepler} quarter, yet no systematic timing error of this kind is known for the \textit{Kepler} space telescope. This makes KPP23 a strong candidate for a new Blazhko RR Lyrae star.  

In contrast, the O$-$C diagram for KPP26 appears to be constant (last panel in Fig.~\ref{fig:cont2}), and no significant frequency can be found in its spectrum, suggesting that its reported amplitude variation is most likely instrumental in origin.

KPP14 is an RRc star for which \citet{Forro2022} did not report amplitude modulation. However, its O$-$C diagram displays a clear, smooth variation (see Fig.~\ref{fig:o-c}). If this variation is periodic, its period may be approximately $\sim982$ days.  

\subsection{Revisited O$-$C spectra}

The Fourier content of the O$-$C diagrams constructed using the template-fitting method differs slightly in some cases from that reported by \citet{Benko2014}. In the case of V783~Cyg, two frequencies were detected ($f_s$ and $f_B - f_s$; see Table~\ref{table:all_freq}) that were not identified by \citet{Benko2014}. The presence of a secondary Blazhko modulation may explain the non-strictly periodic behaviour of the O$-$C curve, which had previously been interpreted as an indication of chaotic modulation \citep{Plachy_2014}.

For V355~Lyr, contrary to the findings of \citet{Benko2014}, the dominant frequency in AM is also dominant in the O$-$C diagram. In the case of V450~Lyr, we detected the same frequencies as \citet{Benko2014}, but with different relative amplitudes. Accordingly, we interpret the signals as $f_B$ and its harmonic 2$f_B$, rather than the sub-harmonic $f_B/2$. This implies that the dominant frequencies differ between AM and FM.  

In the case of V366~Lyr, similarly to V355~Lyr, the dominant frequencies coincide in AM and FM, again in contrast to the results published by \citet{Benko2014}. The difference between the amplitudes of the two frequencies, which determines the order of the dominant and secondary effects, is $4\times10^{-5}$~d, comparable to the uncertainty of the amplitudes ($\sigma \sim 1.6\times10^{-5}$~d).

\subsection{C2C variation on Blazhko stars}

\begin{figure}
    \centering
    \includegraphics[width=0.76\linewidth]{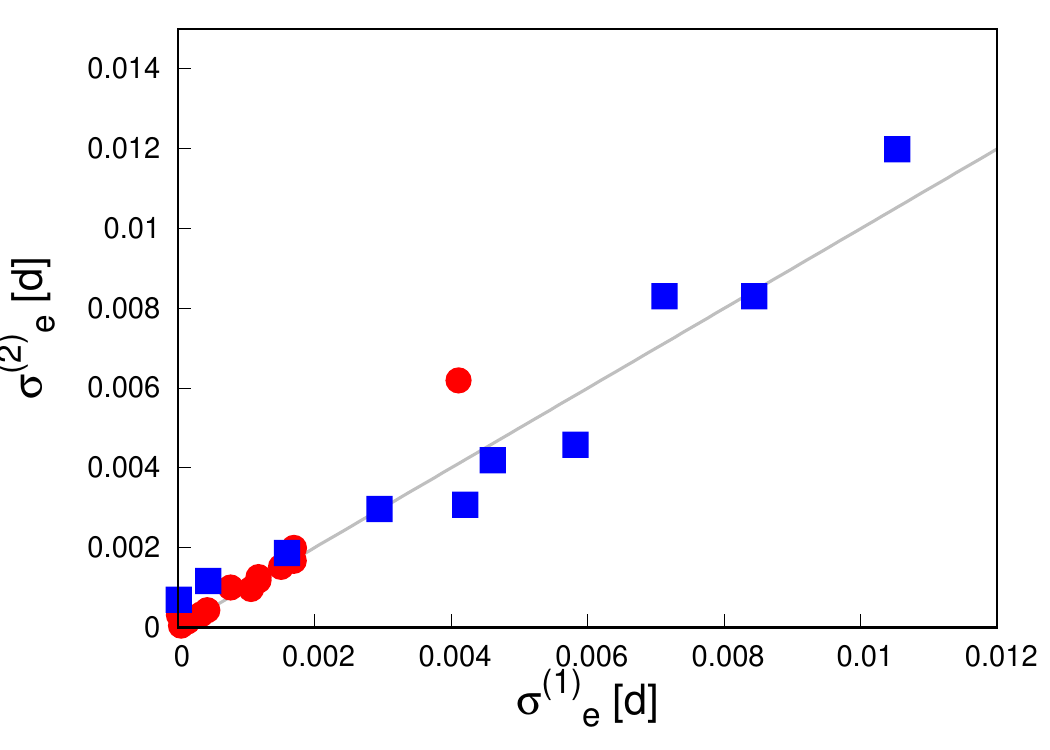}
    \caption{Phase-noise values for the optimal models based on the original O$-$C curves ($\sigma^{(1)}_e$) and the residuals ($\sigma^{(2)}_e$). The gray line with unit slope illustrates the overall similarity of the two sets. Red symbols correspond to \textit{Kepler}'s original target stars, and blue rectangles to stars found in the background pixels.}
    \label{fig:sig_e-sig_e}
\end{figure}

\begin{figure*}
    \centering
    \includegraphics[width=0.34\linewidth]{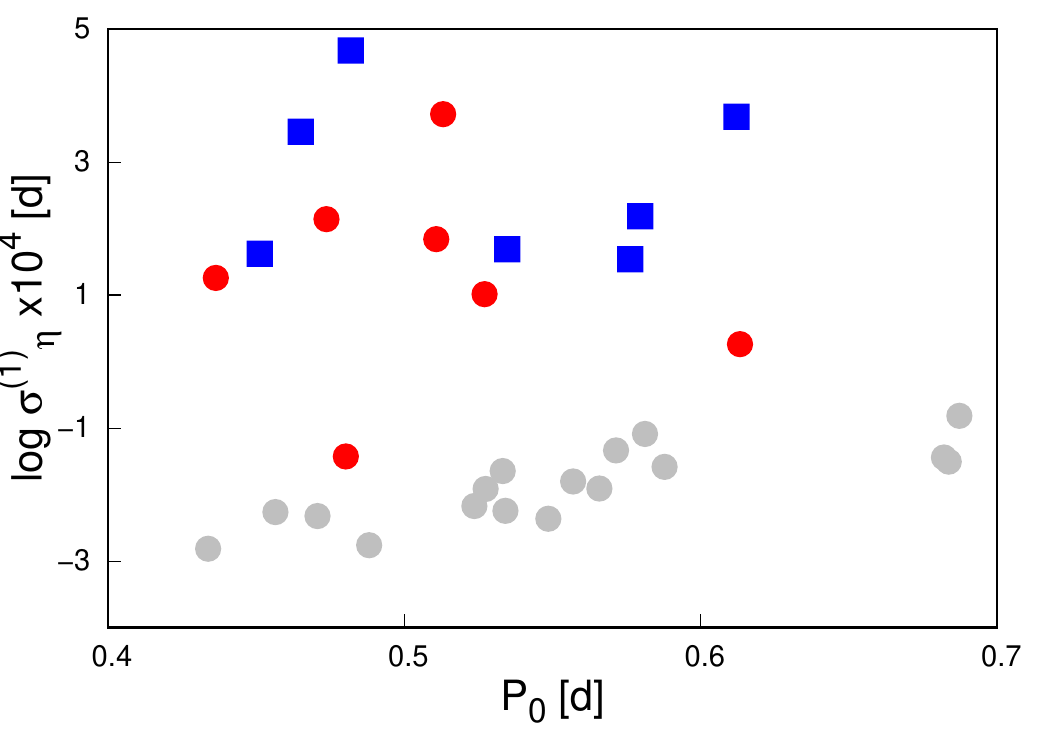}
    \includegraphics[width=0.34\linewidth]{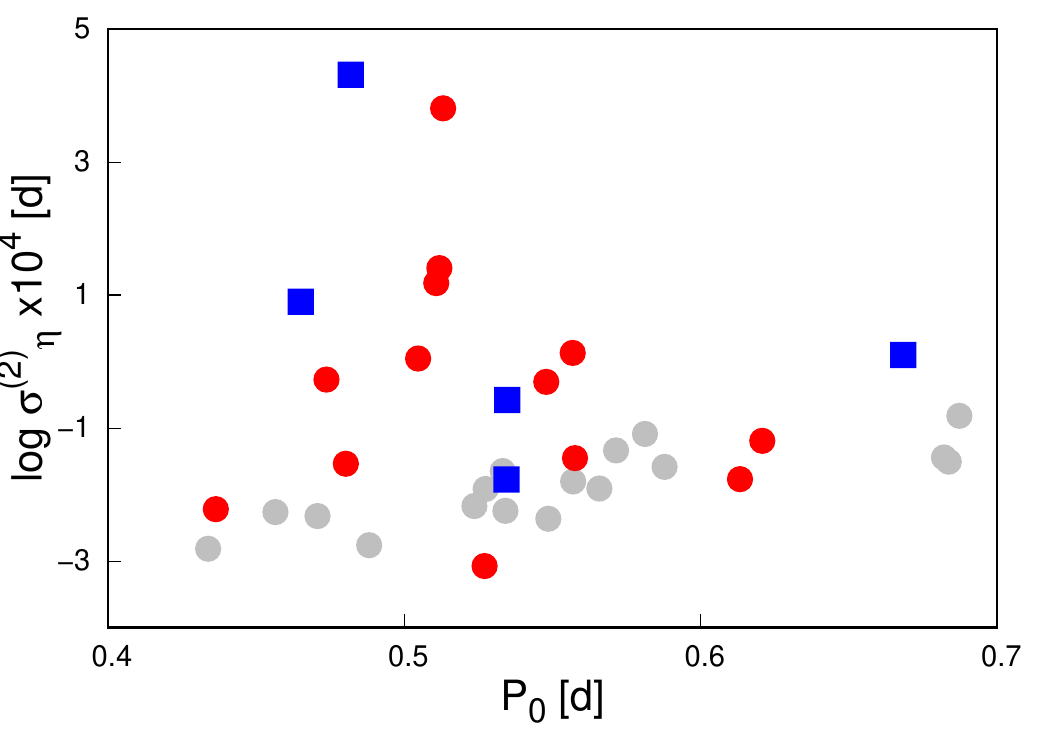}
    \includegraphics[width=0.34\linewidth]{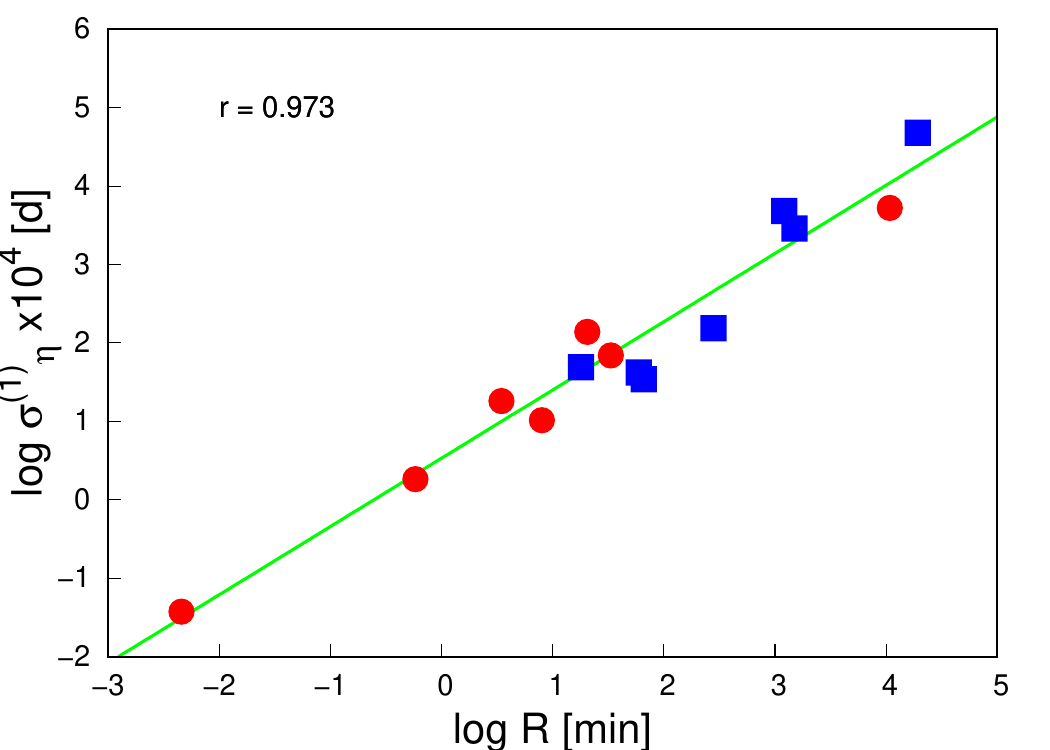}
    \includegraphics[width=0.34\linewidth]{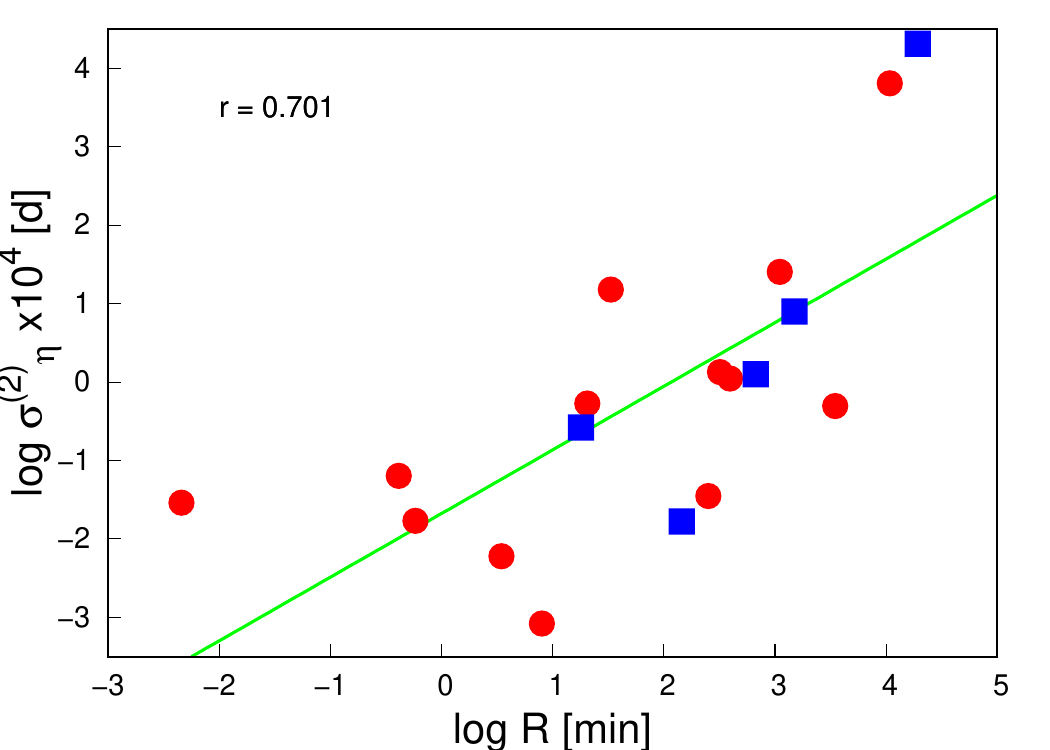}
    \caption{Top: Dependence of the $\sigma_\eta$ parameter characterizing the C2C variation in RRab stars, on the pulsation period for the original O$-$C curves (left) and for the residuals (right). Grey dots represent the sample of non-Blazhko stars from \citet{Benko_2025}. Bottom: $\sigma_\eta$ as a function of the strength of the FM component of the Blazhko effect for the original O$-$Cs (left) and for the residuals (right). Green lines show linear fits; $r$ denotes the Pearson correlation coefficient. Red dots correspond to \textit{Kepler}'s original targets, and blue rectangles to background stars.
    }
    \label{fig:R-sigeta}
\end{figure*}

We applied the same statistical framework as in \citet{Benko_2025}. Here we analysed both the original O$-$C data and the residual O$-$C curves from which all frequencies related to the Blazhko effect had been removed. In contrast to the non-Blazhko stars, the results for certain Blazhko stars were sensitive to the adopted limits of the parameters $\sigma_\mathrm{min}$ and $\sigma_\mathrm{max}$: different bounds often produced significantly different solutions. In such cases, we ran the program 15–20 times with varying limits and selected the most probable global minimum by examining the value of the log-likelihood function.  

Because the phase noise should be identical in both the original and residual O$-$C curves, only solutions for which $\sigma^{(1)}_e$ and $\sigma^{(2)}_e$ are nearly equal were accepted. Here, superscripts (1) and (2) refer to the original and residual curves, respectively. The final accepted values of $\sigma^{(2)}_e$ as a function of $\sigma^{(1)}_e$ are plotted in Fig.~\ref{fig:sig_e-sig_e}. The corresponding $\sigma_e$ values are listed in Columns~8 and~11 of Table~\ref{table:main}. As shown in the figure, $\sigma^{(1)}_e \approx \sigma^{(2)}_e$ holds well. Phase noise tends to be higher for stars detected in background pixels (blue points) than for the original \textit{Kepler} targets (red points), which is expected because the background stars generally have noisier light curves.

The M3 or M4 model proved to be optimal for 24 original RRab O$-$C curves, whereas the M2 model was preferred for KPP01. For 13 RRab stars, the residual O$-$C curves were best described by the M2 model (phase noise plus C2C variation). This outcome is consistent with expectations: once the dominant periodic variations caused by the Blazhko effect are removed, only phase noise and C2C variation should remain. In five cases, the residual curves were still best fitted by the M4 model. This may indicate that the Blazhko-related period variation was not entirely removed, or that a separate long-term period variation is present. The same explanation applies to the four stars whose residuals were dominated by real period variations (M3). For the three stars where the residuals were best fitted by the M1 model, the high noise level likely masked any detectable C2C variation.  No residuals were constructed for the two RRc stars (KPP14 and KPP15), their original O$-$C diagrams are described by the M2 and M4 models.

The parameters $\sigma^{(1)}_e$, $\sigma^{(1)}_\eta$, and $\sigma^{(1)}_\xi$ obtained from the optimal models of the original O$-$C diagrams are listed in Columns~8–10 of Table~\ref{table:main}, while the corresponding values for the residuals, $\sigma^{(2)}_e$, $\sigma^{(2)}_\eta$, and $\sigma^{(2)}_\xi$, are given in Columns~11–13. The IDs of the optimal models for both the original and residual O$-$C diagrams are shown in Column~14.  

As \citet{Benko_2025} found a correlation between the strength of the C2C variation ($\sigma_\eta$) and the pulsation period for non-Blazhko stars, we examined whether a similar relationship exists for Blazhko stars. The upper-left panel of Fig.~\ref{fig:R-sigeta} presents the $\sigma^{(1)}_\eta$ values obtained from the original O$-$C curves of RRab stars. Due to the large range of values, a logarithmic vertical scale was used. 
The comparison between Blazhko and non-Blazhko stars shows that (i) The $\sigma^{(1)}_\eta$ values are significantly higher than those of the non-Blazhko RRab stars (grey dots in the figure), and (ii) no clear dependence on period is apparent.  

The upper-right panel of Fig.~\ref{fig:R-sigeta} shows the period dependence of the $\sigma^{(2)}_\eta$ values obtained from the residual O$-$C diagrams. Although some values fall within the range observed for non-Blazhko stars, the mean value, $\langle \sigma^{(2)}_\eta \rangle = 7.18\times10^{-4}$, remains higher than that for non-Blazhko RRab stars ($\langle \sigma_\eta(\mathrm{RRab}) \rangle = 1.76\times10^{-5}$) and RRc stars ($\langle \sigma_\eta(\mathrm{RRc}) \rangle = 1.20\times10^{-4}$). No evident period dependence is observed in this case either.  

To quantify the strength of the FM component of the Blazhko effect as reflected in the O$-$C diagrams, we introduced the parameter
\begin{equation}
    R = \sqrt{\sum_{i=1}^n A_i^2} ,
\end{equation}
where $A_i$ are the amplitudes (in minutes) of the frequencies associated with the Blazhko effect in the O$-$C spectrum. $R$ serves as a global index of the strength and complexity of the FM modulation, since it increases when additional harmonics or combination peaks are present or when their amplitudes are larger.

The lower-left panel of Fig.~\ref{fig:R-sigeta} shows $\sigma^{(1)}_\eta$, which characterises the strength of the C2C variation, plotted against the $R$ parameter representing the strength of the Blazhko FM component. A very strong correlation is evident, with a Pearson correlation coefficient of $r = 0.973$. A similar plot for the residuals (lower-right panel) reveals a slightly weaker but still strong correlation ($r = 0.701$). This effect likely reflects that part of the Blazhko variation is absorbed into the $\sigma^{(1)}_\eta$ term when fitting the original O$-$C curves, whereas the residuals isolate the intrinsic C2C component.

This finding suggests that the C2C variation and the Blazhko effect are interrelated, although the physical nature of this relationship is unclear. At this stage, it remains unclear whether the Blazhko effect amplifies the C2C variation, whether strong C2C variability contributes to the occurrence of the Blazhko effect, or whether both phenomena arise from a common underlying cause — or even coincide by chance. A definitive conclusion would require a theoretical explanation both for the origin of the C2C variation and for the mechanism of the Blazhko effect, which is currently lacking.

\section{Conclusions}

In this Letter, we investigated the C2C variations of RR~Lyrae stars showing the Blazhko effect using the same statistical approach previously applied to non-Blazhko stars \citep{Benko_2025}. \citet{Benko_2025} demonstrated that the irregular O$-$C variations observed in most non-Blazhko RR~Lyrae stars can be explained as the consequence of random C2C fluctuations in their light curves.  

(i) We arrived at a similar conclusion here, finding that the M2 or M4 model (both incorporating C2C variation) provides the optimal description for the O$-$C curves of 20 stars in our 27-object sample. 
The parameter $\sigma_\eta$, which characterises the strength of the C2C variation, is on average larger for Blazhko stars than for both non-Blazhko RRab and RRc stars. This result qualitatively explains why Blazhko stars tend to exhibit stronger irregular O$-$C variations than non-Blazhko ones.

(ii) The strength of the FM component of the Blazhko effect shows a clear positive correlation with the C2C variation strength:  stronger Blazhko effect corresponds to a larger $\sigma_\eta$ value.

(iii) As a by-product of our analysis, we identified two background RR~Lyrae stars (KPP14 and KPP23) in the original \textit{Kepler} field showing clear phase variations, and we propose them as new Blazhko candidates.

\begin{acknowledgements}
This paper includes data collected by the \textit{Kepler} mission. Funding for the mission is provided by the NASA Science Mission Directorate. The research was partially supported by the ‘SeismoLab’ KKP-137523 \'Elvonal grant of the Hungarian Research, Development and Innovation Office (NKFIH) 
Some {\sc python} codes were developed with the help of ChatGPT 5.0 
\end{acknowledgements}

\bibliographystyle{aa}
\bibliography{oc_Blazhko.bib}

\begin{appendix}

\onecolumn
\captionsetup[subfigure]{labelformat=empty}

\begin{figure*}[ht]
\section{O$-$C diagrams of Blazhko RR Lyrae stars in the original \textit{Kepler} field.}
\subfloat[]{\includegraphics[width=0.9\textwidth]{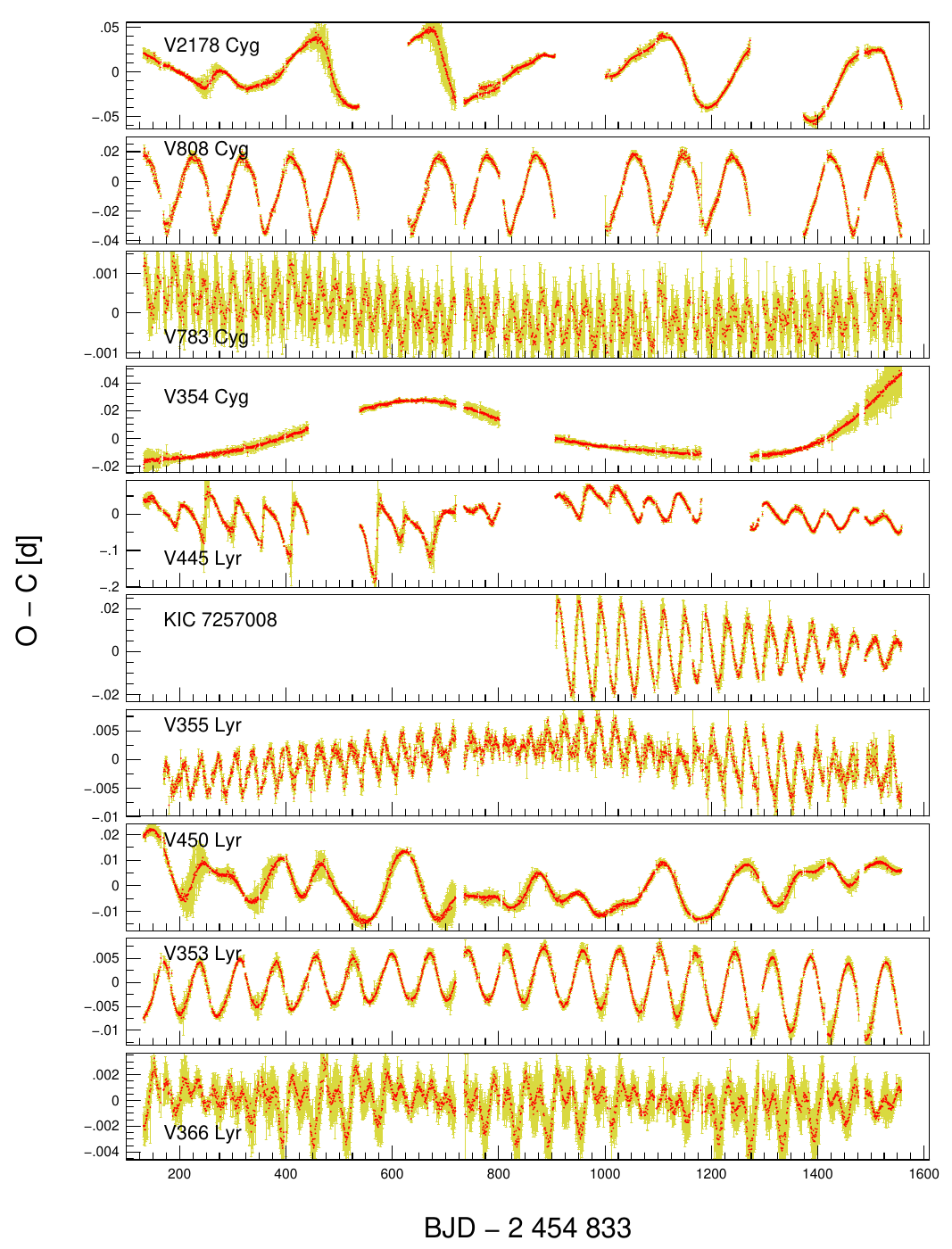}}
\caption{
O$-$C diagrams of \textit{Kepler} Blazhko RR Lyrae stars prepared using the template fitting method. 
Red curves: the main target stars of the \textit{Kepler} mission, for which the light curves were taken from 
\citet{Benko2014}; blue curves: stars found by \citet{Forro2022} in the background pixels.
There can be a difference of four orders of magnitude in the O$-$C values of individual stars (see vertical scales).
The light yellow bars show errors estimated based on Monte Carlo simulation.
}\label{fig:o-c}
\end{figure*}
\begin{figure*}
\ContinuedFloat
\subfloat[]{\includegraphics[width=0.9\textwidth]{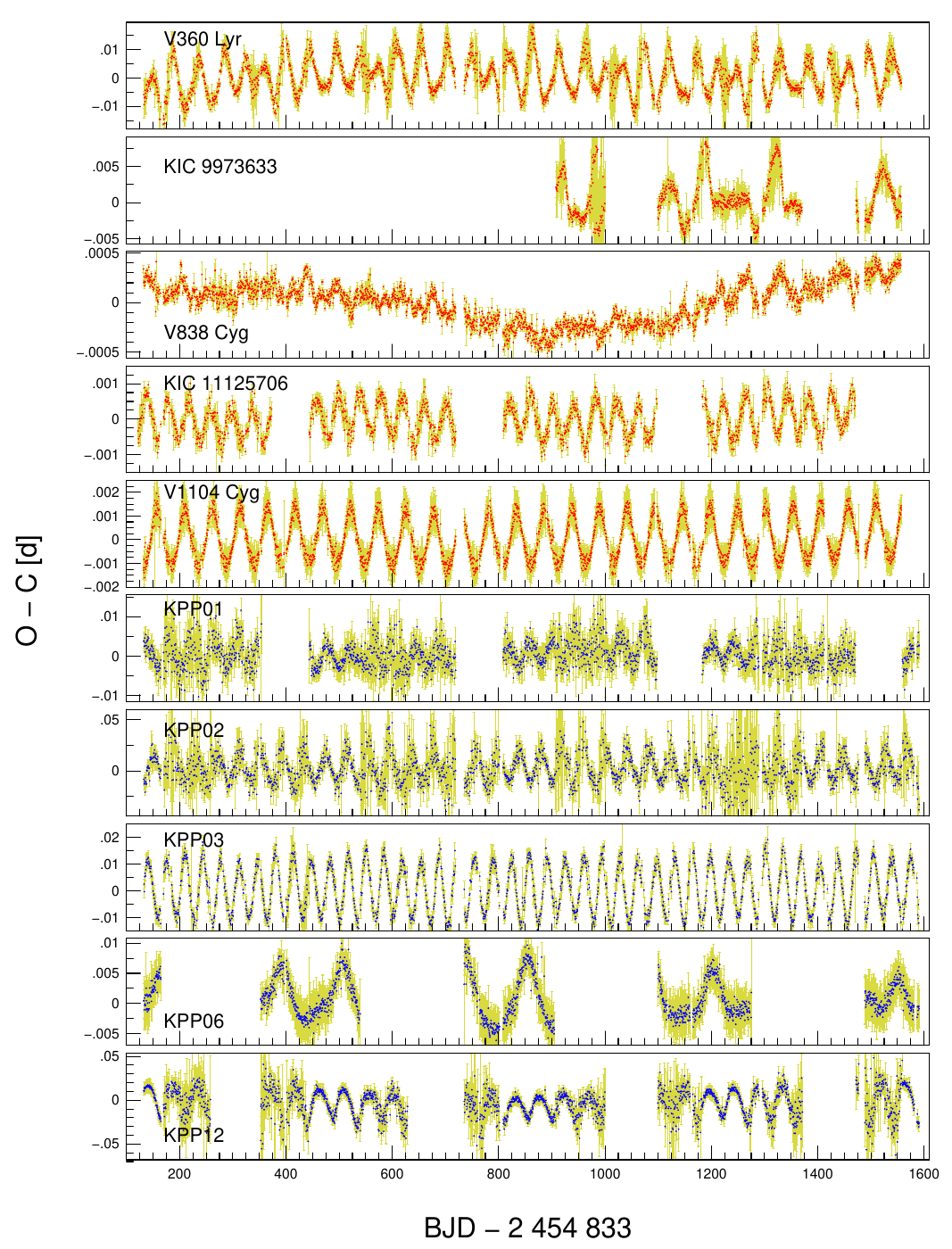}}
\caption{continued}
\label{fig:cont1}
\end{figure*}
\begin{figure*}
\ContinuedFloat
\subfloat[]{\includegraphics[width=0.9\textwidth]{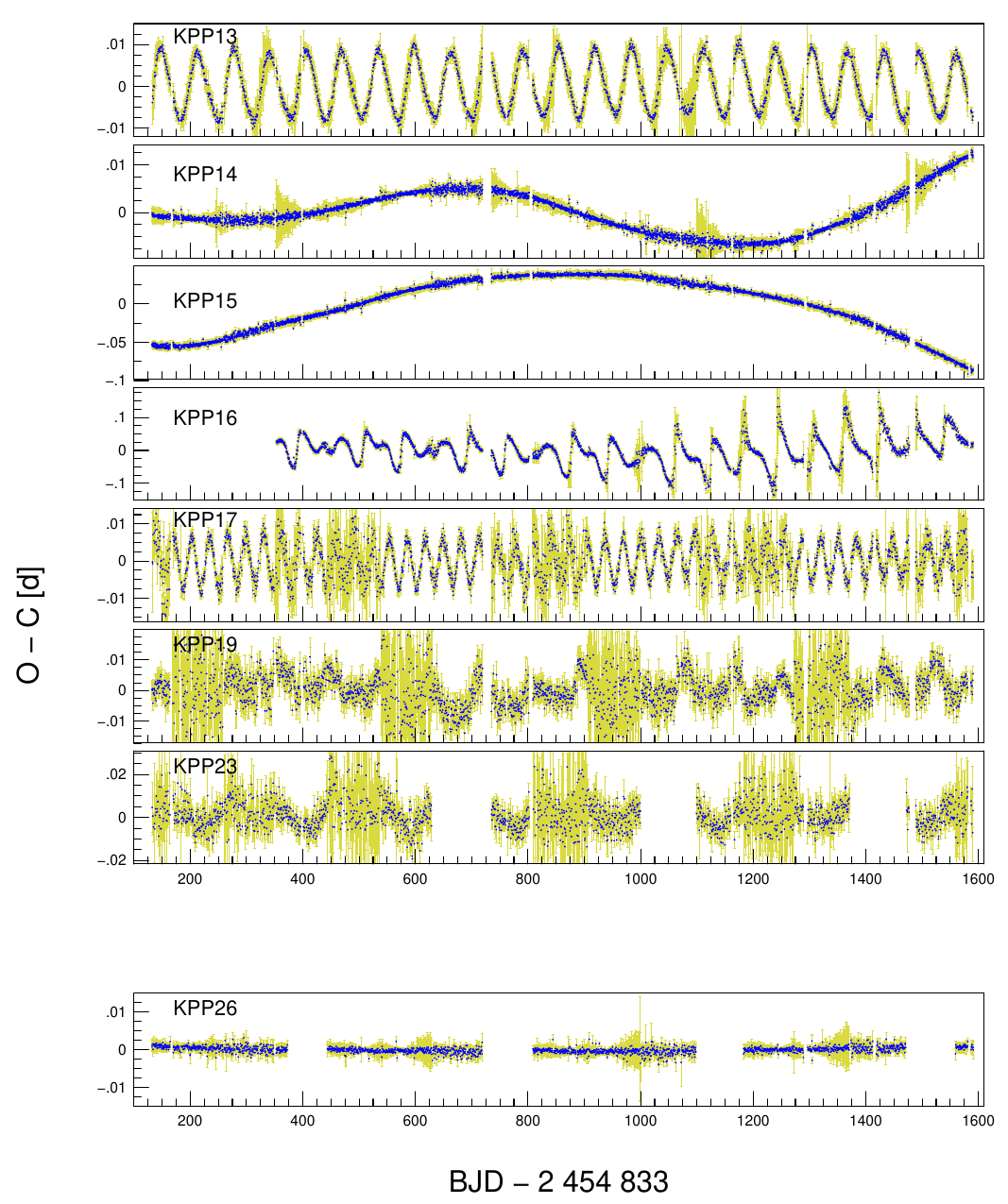}}
\caption{continued. The last panel shows KPP26, in which \citet{Forro2022} found uncertain amplitude variations in addition to KPP23.}
\label{fig:cont2}
\end{figure*}

%
%

\begin{table*}[htb]
\caption{Pre-whitened frequencies of the O$-$C diagrams}\label{table:all_freq}
\small
\centering

\begin{minipage}[t]{0.48\textwidth}
\vspace{0pt}
\centering
\begin{tabular}{llllL}
\hline\hline
\noalign{\smallskip}
Star & Frequency & Amplitude & Phase & \mathrm{Identification} \\
     & d$^{-1}$  & d         & rad   & \\
\noalign{\smallskip}
\hline
\noalign{\smallskip}
V2178 Cyg	&	 0.0045855	&	   0.02812	&	     1.794	&	f_B		\\	
 	&	 0.0091711	&	   0.00504	&	     6.130	&	2f_B		\\	
 	&	 0.0060212	&	   0.01481	&	     2.662	&	f_s		\\	
V808 Cyg	&	 0.0108539	&	   0.02360	&	     4.982	&	f_B		\\	
 	&	 0.0217077	&	   0.00309	&	     5.896	&	2f_B		\\	
 	&	 0.0325616	&	   0.00275	&	     0.627	&	3f_B		\\	
 	&	 0.0434154	&	   0.00076	&	     1.149	&	4f_B		\\	
V783 Cyg	&	 0.0361554	&	   0.00046	&	     2.398	&	f_B		\\	
 	&	 0.0723108	&	   0.00006	&	     3.205	&	2f_B		\\	
 	&	 0.0258845	&	   0.00008	&	     3.470	&	f_s		\\	
 	&	 0.0102708	&	   0.00003	&	     4.974	&	f_B-f_s		\\	
V354 Lyr	&	 0.0010077	&	   0.02088	&	     3.752	&	f_B		\\	
 	&	 0.0020154	&	   0.00544	&	     0.326	&	2f_B		\\	
V445 Lyr	&	 0.0181799	&	   0.03439	&	     3.331	&	f_B		\\	
 	&	 0.0363597	&	   0.00839	&	     0.759	&	2f_B		\\	
 	&	 0.0069436	&	   0.00955	&	     0.397	&	f_s		\\	
 	&	 0.0197406	&	   0.00989	&	     5.135	&	f_t		\\	
 	&	 0.0379205	&	   0.00752	&	     2.414	&	f_B+f_t		\\	
 	&	 0.0251235	&	   0.00618	&	     4.407	&	f_B+f_s		\\	
KIC 7257008	&	 0.0254090	&	   0.01354	&	     6.072	&	f_B		\\	
 	&	 0.0242706	&	   0.00311	&	     1.085	&	f_B-f_s		\\	
 	&	 0.0262676	&	   0.00323	&	     1.328	&	f_B+f_s		\\	
 	&	 0.0496933	&	   0.00181	&	     0.498	&	2f_B-f_s		\\	
 	&	 0.0509985	&	   0.00188	&	     5.056	&	2f_B		\\	
 	&	 0.0124425	&	   0.00149	&	     4.168	&	f_B/2		\\	
V355 Lyr	&	 0.0322772	&	   0.00189	&	     1.993	&	f_B		\\	
 	&	 0.0307753	&	   0.00145	&	     0.327	&	f_B-f_\mathrm{trend}		\\	
 	&	 0.0616018	&	   0.00083	&	     1.644	&	f_s		\\	
 	&	 0.0646639	&	   0.00028	&	     3.442	&	f_s+f_\mathrm{trend}		\\	
 	&	 0.1632057	&	   0.00040	&	     4.727	&	f_t(?)		\\	
 	&	 0.0938791	&	   0.00020	&	     4.631	&	f_B+f_s		\\	
V450 Lyr	&	 0.0062374	&	   0.00556	&	     2.578	&	f_B		\\	
 	&	 0.0080658	&	   0.00457	&	     1.204	&	f_s		\\	
 	&	 0.0044090	&	   0.00477	&	     2.840	&	2f_B-f_s		\\
    &	 0.0124749	&	   0.00277	&	     2.311	&	2f_B		\\
 	&	 0.0088181	&	   0.00210	&	     5.461	&	2f_B-2f_s		\\	
V353 Lyr	&	 0.0140205	&	   0.00790	&	     5.564	&	f_B		\\	
 	&	 0.0003910	&	   0.00393	&	     5.946	&	f_\mathrm{trend}(?)		\\	
 	&	 0.0142522	&	   0.00302	&	     1.712	&	2f_s		\\	
 	&	 0.0075273	&	   0.00067	&	     4.716	&	f_s		\\	
 	&	 0.0203894	&	   0.00046	&	     3.667	&	f_B+f_s		\\	
 	&	 0.0277425	&	   0.00036	&	     2.895	&	?		\\	
 	&	 0.0283561	&	   0.00025	&	     2.266	&	2f_B		\\	
V366 Lyr	&	 0.0341299	&	   0.00105	&	     1.396	&	f_B		\\	
 	&	 0.0158951	&	   0.00101	&	     3.799	&	f_s		\\	
 	&	 0.0182338	&	   0.00065	&	     3.145	&	f_B-f_s		\\	
 	&	 0.0317581	&	   0.00052	&	     2.129	&	2f_s		\\	
 	&	 0.0023489	&	   0.00037	&	     4.942	&	f_B-2f_s		\\	
V360 Lyr	&	 0.0237200	&	   0.00498	&	     4.045	&	f_s/2		\\	
 	&	 0.0191135	&	   0.00475	&	     4.659	&	f_B		\\	
 	&	 0.0282590	&	   0.00295	&	     6.232	&	f_s-f_B		\\	
 	&	 0.0383137	&	   0.00087	&	     0.988	&	2f_B		\\	
 	&	 0.0090008	&	   0.00083	&	     3.237	&	f_s-2f_B		\\	
 	&	 0.0474543	&	   0.00075	&	     4.922	&	f_s		\\	
 	&	 0.0430165	&	   0.00068	&	     0.989	&	f_s/2+f_B		\\	
 	&	 0.0666808	&	   0.00026	&	     2.898	&	f_s+f_B		\\	
KIC 9973633	&	 0.0148140	&	   0.00295	&	     4.274	&	f_B		\\	
 	&	 0.0296279	&	   0.00073	&	     0.485	&	2f_B		\\	
 	&	 0.0374844	&	   0.00082	&	     4.014	&	f_s		\\	
 	&	 0.0517588	&	   0.00051	&	     2.683	&	f_B+f_s		\\	
V838 Cyg	&	 0.0169100	&	   0.00007	&	     5.435	&	f_B		\\	
KIC 11125706	&	 0.0248450	&	   0.00055	&	     5.265	&	f_B		\\	
V1104 Cyg	&	 0.0192344	&	   0.00118	&	     1.308	&	f_B		\\	
 	&	 0.0384688	&	   0.00012	&	     1.256	&	2f_B		\\	
KPP01	&	 0.0235854	&	   0.00244	&	     6.019	&	f_B		\\	

\end{tabular}
\end{minipage}%
\hfill
\begin{minipage}[t]{0.48\textwidth}
\vspace{0pt}
\centering
\begin{tabular}{llllL}
\hline\hline
\noalign{\smallskip}
Star & Frequency & Amplitude & Phase & \mathrm{Identification} \\
     & d$^{-1}$  & d         & rad   & \\
\noalign{\smallskip}
\hline
\noalign{\smallskip}
KPP02	&	 0.0246673	&	   0.01515	&	     3.354	&	f_B		\\	
KPP03	&	 0.0293182	&	   0.01163	&	     0.549	&	f_B		\\	
 	&	 0.0357913	&	   0.00151	&	     3.462	&	f_s		\\	
 	&	 0.0586364	&	   0.00051	&	     0.934	&	2f_B		\\	
KPP06	&	 0.0086548	&	   0.00346	&	     5.404	&	f_B		\\	
 	&	 0.0173095	&	   0.00175	&	     2.477	&	2f_B		\\	
 	&	 0.0080142	&	   0.00133	&	     2.370	&	f_s		\\	
KPP10	&	 0.0099419	&	   0.00092	&	     0.704	&	f_B		\\	
KPP12	&	 0.0217491	&	   0.01628	&	     1.169	&	f_B		\\	
 	&	 0.0434983	&	   0.00309	&	     3.336	&	2f_B		\\	
 	&	 0.0327370	&	   0.00169	&	     3.008	&	f_s		\\	
KPP13	&	 0.0155897	&	   0.00795	&	     5.642	&	f_B		\\	
 	&	 0.0311794	&	   0.00100	&	     4.582	&	2f_B		\\	 
    &	 0.0224442	&	   0.00055	&	     5.864	&	f_s		\\	
KPP16	&	 0.0164988	&	   0.03745	&	     3.359	&	f_B		\\	
 	&	 0.0329976	&	   0.01581	&	     0.351	&	2f_B		\\	
 	&	 0.0494964	&	   0.00679	&	     3.577	&	3f_B		\\	
 	&	 0.0271194	&	   0.02411	&	     2.182	&	f_s		\\	
 	&	 0.0436182	&	   0.00946	&	     5.661	&	f_B+f_s		\\	
 	&	 0.0106207	&	   0.01063	&	     4.554	&	f_s-f_B		\\	
 	&	 0.0601170	&	   0.00561	&	     2.503	&	2f_B+f_s		\\	
KPP17	&	 0.0313476	&	   0.00603	&	     5.491	&	f_B		\\	
KPP19	&	 0.0111540	&	   0.00341	&	     1.499	&	f_B		\\	
 	&	 0.0223080	&	   0.00188	&	     1.257	&	2f_B		\\	
KPP23	&	 0.0054306	&	   0.00144	&	     3.658	&	f_B		\\	
 	&	 0.0108613	&	   0.00403	&	     2.229	&	2f_B		\\	
\end{tabular}
\end{minipage}

\end{table*}

\begin{sidewaystable*}
    \caption{\textit{Kepler} Blazhko RR Lyrae stars, to whose O$-$C diagrams we fitted statistical models.
    }\label{table:main}
    \smallskip
    {\scriptsize
    \begin{tabular}{lll S[table-format=4.3] l S[table-format=3.2] l S[table-format=2.2] S[table-format=2.2] l S[table-format=3.2]  S[table-format=2.2]ll}
    \hline\hline                 
    \noalign{\smallskip}
     Name & Alternate ID & $P_0$ & $P_{\mathrm B}$ & $\sigma(P_{\mathrm B})$ & $R$ & $\sigma(R)$ & {$\sigma^{(1)}_e$} & {$\sigma^{(1)}_\eta$} & {$\sigma^{(1)}_\xi$} & {$\sigma^{(2)}_e$} & {$\sigma^{(2)}_\eta$} &  {$\sigma^{(2)}_\xi$} & Model\\    
        &    &   {(d)} & {(d)} & {(d)} & {(min)} & {(min)} & {($\times10^{-4}$d)} & {($\times10^{-4}$d)}& {(d)} & {($\times10^{-4}$d)} & {($\times10^{-4}$d)}& {(d)} & \\
        \noalign{\smallskip}
    \hline                        
    \noalign{\smallskip}
         KPP14		&	ATO J283.5884+43.0995		&	0.309402581	&	982	&	3	&	7.79	&	0.07	&	7.28 & 	0.21 &	 5.68$\times10^{-7}$ & & & & M4 \\
KPP15		&	ZTF J192145.16+430142.4 	&	0.35290612	&	2496	&		&	126.8	&	0.2	&	18.44	& 5.07	&	0 & & & & M2 \\
V1104 Cyg	&	KIC 12155928			&	0.4363851	&	51.990	&	0.005	&	1.715 	&	0.007	&	1.33	&	3.52	&	1.29$\times10^{-10}$  &2.01 &0.11 &0 & M4->M2 \\
KPP06		&	ZTF J195301.70+403852.6		&	0.451216965	&	115.5	&	0.2	&	5.9	&	0.3	&	4.41 	&	5.06 	&	1.66$\times10^{-15}$ & 11.63& 0& 0& M4->M1 \\
KPP12		&	ZTF J193816.70+415811.2		&	0.46509111	&	45.98	&	0.03	&	24.0 	&	0.8	&	71.27 	&	31.73 	&	4.67$\times10^{-9}$ & 83.02 & 2.46 & 0 & M4->M2 \\
V355 Lyr	&	KIC 7505345			&	0.473699	&	30.99	&	0.02	&	3.7	&	0.2	&	10.72 	&	8.51 	&	1.96$\times10^{-7}$  &9.64 & 0.76& 1.00$\times10^{-10}$& M4->M4 \\
V838 Cyg	&	KIC 10789273			&	0.480280	&	59.1	&	0.2	&	0.096	&	0.01	&	0.45 	&	0.24 	&	6.57$\times10^{-11}$ &0.48 & 0.22& 0 & M4->M2 \\
KPP16		&	ZTF J184738.98+434147.2		&	0.48195322	&	60.49	&	0.03	&	72	&	1	&	46.11	&	107.8	&	1.00$\times10^{-7}$  & 41.96 & 74.81& 1.00$\times10^{-5}$ & M4->M4 \\
V2178 Cyg	&	KIC 3864443			&	0.4869470	&	218.1	&	0.3	&	46.3	&	0.6	&	15.09 	&	0 	&	5.47$\times10^{-5}$ & 15.18& 0& 4.66$\times10^{-5}$& M3->M3 \\
V450 Lyr	&	KIC 7671081			&	0.5046198	&	160.3	&	0.2	&	13.4	&	0.2	&	4.32 	&	0 	&	2.53$\times10^{-5}$ & 4.31 & 1.05 & 1.50$\times10^{-5}$& M3->M4 \\
KIC 9973633	&	ATO J299.7044+46.8491		&	0.510783	&	67.5	&	0.1	&	4.6	&	0.1	&	7.73 	&	6.29 	&	1.39$\times10^{-9}$ & 10.03& 3.25&0 & M4->M2 \\
KIC 7257008	&	ATO J281.8641+42.8313		&	0.511787	&	39.36	&	0.02	&	21	&	1	&	11.79 	&	0 	&	2.82$\times10^{-4}$ & 12.64& 4.08& 0& M3->M2 \\
V445 Lyr	&	KIC 6186029			&	0.5130907	&	55.01	&	0.04	&	56	&	2	&	41.1 	& 41.34	 & 2.09$\times10^{-3}$ & 61.94& 45.16& 0 & M4->M2 \\
V366 Lyr	&	KIC 9578833			&	0.5270284	&	29.300	&	0.005	&	2.47 	&	0.02	&	0.14 	&	2.75 	&	1.00$\times10^{-11}$ &3.14 & 0.05& 0& M4->M2 \\
KPP19		&	ATO J294.1627+45.0491		&	0.529213455	&	89.65	&	0.1	&	5.6	&	0.3	&	58.22	&	0	&	9.99$\times10^{-6}$  & 45.83& 0& 9.48$\times10^{-6}$& M3->M3 \\
KPP17		&	ATO J296.1407+43.6197		&	0.534414036	&	31.900	&	0.006	&	8.7	&	0.1	&	42.07	&	0	&	2.62$\times10^{-4}$ & 30.74 & 0.17 & 1.21$\times10^{-10}$ & M3->M4 \\
KPP01		&	ATO J292.8211+37.6989		&	0.534705027	&	42.40	&	0.05	&	3.5	&	0.2	&	29.50	&	5.42 	& 0	 & 29.74 & 0.56 & 0 & M2->M2 \\
V808 Cyg	&	KIC 4484128			&	0.5478635	&	92.13	&	0.01	&	34.51	&	0.09	&	16.98	&	0	&	1.32$\times10^{-4}$ & 16.72& 0.74& 0 & M3->M2 \\
V353 Lyr	&	KIC 9001926			&	0.5567997	&	71.3	&	0.3	&	12	&	2	&	3.35 	& 0	&	4.22$\times10^{-5}$ & 3.25 & 1.14& 0& M3->M2 \\
V360 Lyr	&	KIC 9697825			&	0.5575755	&	52.3	&	0.1	&	11	&	1.5	&	16.98 	&	0 	&	1.99$\times10^{-4}$  & 19.94& 0.23&2.87$\times10^{-7}$ & M3->M4 \\
V354 Lyr	&	KIC 6183128			&	0.5616892	&	892	&	2	&	31.1	&	0.2	&	11.82 	&	0 	&	3.09$\times10^{-6}$  & 11.82& 0& 2.73$\times10^{-6}$& M3->M3 \\
KPP23		&	ATO J297.7643+47.9708		&	0.576200749	&	184	&	1.6	&	6.2	&	0.6	&	84.4	&	4.64	&	7.63$\times10^{-8}$ & 83.1 & 0 & 0 & M4->M1 \\
KPP13		&	ZTF J192354.75+423607.8		&	0.579570774	&	61.145	&	0.006	&	11.57	&	0.04	&	0.13	&	8.89	&	1.22$\times10^{-4}$ &6.96 & 0 & 5.33$\times10^{-8}$& M4->M3 \\
KPP02		&	ZTF J190247.20+380429.3		&	0.61204451	&	40.54	&	0.017	&	21.8	&	0.5	&	105.34	&	39.73	&	1.00$\times10^{-10}$ & 119.92 & 0 & 0 & M4->M1 \\
KIC 11125706	&	ROTSE1 J190058.77+484441.5	&	0.6132200	&	40.25	&	0.01	&	0.79	&	0.01	&	1.07 	&	1.30 	&	1.03$\times10^{-9}$ &1.85 &0.17 &0 & M4->M2 \\
V783 Cyg	&	KIC 5559631			&	0.6207001	&	27.658	&	0.007	&	0.68	&	0.02	&	1.40	&	0 	&	2.40$\times10^{-5}$ &1.50 & 0.30&0 & M3->M2 \\
KPP03		&	Gaia DR3 2052744366034050048	&	0.668340314	&	31.108	&	0.003	&	16.9	&	0.1	&	15.96 	&	0 	&	4.29$\times10^{-4}$ & 18.70 & 1.11 & 0 & M3->M2 \\
\hline

    \end{tabular}
    }
    \tablefoot{Column 1: short name used in this work; column 2: alternate name; column 3: pulsation period from \citet{Benko2014} and \citet{Forro2022}; columns 4-5: Blazhko period and its accuracy corresponding to the frequency with the highest amplitude in the O$-$C spectra; columns 6-7: the Blazhko FM strength parameter $R$ and its calculated error; 
    columns 8–10 and 11–13: parameter values of the optimal statistical model fitted to the original and residual O-C curves, respectively; column 14: model identifiers for the original and residual O$-$Cs (see text for the details).}
    
\end{sidewaystable*}

\end{appendix}

\end{document}